\documentclass[preprint,aps,nofootinbib]{revtex4}
\usepackage{}
\usepackage{graphicx}
\usepackage{amsmath}
\usepackage{amsfonts}
\usepackage{mathrsfs}
\usepackage{amssymb}
\usepackage{color}%
\usepackage{dcolumn}
\providecommand{\U}[1]{\protect\rule{.1in}{.1in}}

\newcommand{\f}{\begin{equation}}
\newcommand{\ff}{\end{equation}}
\newcommand{\fa}{\begin{eqnarray}}
\newcommand{\ffa}{\end{eqnarray}}

\begin{document}

\title{Evolution of interior entropy of a loop quantum-corrected black hole pierced by a cosmic string}

\author{Wei Zhang $^{1,2}$}
\email{w.zhang@mail.bnu.edu.cn}

\affiliation{$^1$School of Physics and Astronomy, Beijing Normal University, Beijing 100875, China\\
$^2$Key Laboratory of Multiscale Spin Physics, Ministry of Education, Beijing Normal University, Beijing 100875, China}

\begin{abstract}
The concept of interior entropy of black hole provides a possible way to deal with information problem. We derive evolution relation between interior entropy and Bekenstein-Hawking entropy for a loop quantum-corrected black hole with large mass pierced by an infinitely straight cosmic string during Hawking radiation. 
We find that the present cosmic string can influence the evolution relation between these two types of entropy under black hole evaporation 
and the second law of thermodynamics is satisfied for the total variation
of these two types of entropy. 
\end{abstract}
\maketitle

\section {Introduction}
We know that information problem is very important in the field of
black hole thermodynamics and many schemes so far have been proposed
to resolve it. The concept of interior entropy
of black hole provides a choice to deal with this
issue. Specifically,
the authors in \cite{Christodoulou:2014yia} start from the calculation 
of volume of flat spacetime and find a way to compute the
interior volume of static spherically symmetric black hole related to curved spacetime. They adopt advanced
Eddington-Finkelstein coordinate $(v, r, \theta, \phi)$ to express the metric and employ the Lagrangian method to
get the formula of interior volume in terms of static spherically symmetric black hole case as
\begin{equation}
V_{\Sigma}=
\left[\sqrt{-r^4 f(r)}\right]_{max}\int_{\Omega}\sin\theta d{\theta}d{\phi}\int_{0}^{v}dv
=3\sqrt{3}\pi M^2 v, \label{aa}
\end{equation}
where $\lbrack \rbrack_{max}$ stands for the extreme value wherein, $v$ denotes the advanced Eddington time, and
the initial time of the lower integration bound
has been set to be zero.
Soon after, \cite{Zhang:2015gda} considers a massless scalar field distributing in the interior volume of
black hole and uses quantum statistical method to calculate the corresponding interior entropy
$S_\Sigma$ which is proportion to the interior volume yields
\begin{equation}
S_\Sigma=\frac{\pi^2 T_H^3}{45}V_\Sigma, \label{ab}
\end{equation}
with $T_H$ the Hawking temperature of the spherically symmetric black hole.
Up to now, a series of works \cite{Bengtsson:2015zda,Ong:2015tua,Ong:2015dja,Ong:2016xcq,Astuti:2016dmk,Christodoulou:2016tuu,Bhaumik:2016sav,
Zhang:2016sjy,Wang:2017zfn,Gusin:2017zle,Wang:2018dvo,Yang:2018arj,Han:2018jnf,Wang:2018txl,Ali:2018sqk,
Wang:2019ake,Zhang:2019pzd,Wang:2019dpk,Wang:2019ear,Ali:2019icq,Zhang:2019abv,Ali:2020olc,Chew:2020twk,Wang:2020fgz,
Jiang:2020rxx,Wen:2020thi,Maurya:2022vjd}
have been done concern with interior volume and entropy in different black hole models   
and a review article \cite{Ali:2024txh} refers to this topic.
All of these works neglect the effects of quantum gravity
except \cite{Zhang:2019abv} which introduced it through correcting
Bekenstein-Hawking entropy and temperature of black hole in such an 
indirect manner. 

On the other hand, the authors of \cite{Lewandowski:2022zce} 
apply loop quantum gravity (LQG) method
and obtain a loop quantum-corrected Schwarzschild black
hole who contains quantum gravity effects in the metric directly via quantum Oppenheimer-Snyder model. 
The effects of quantum gravity of the obtained black hole could hide in the observable hole's shadow\cite{Yang:2022btw}. 
Besides, the hole's singularity will be removed and there exists a minimal mass of the hole at Planck scale which means
that there will be a stable
remnant at the end of Hawking radiation for the black hole.
Furthermore,
\cite{Han:2023wxg} finds a geometry of black to white hole transition 
for the loop quantum-corrected Schwarzschild
black hole and \cite{Han:2024rqb} further calculates the spin foam amplitude of this quantum tunneling transition in LQG.
These works implying that the black hole information may be eventually either
store in a remnant or release through a white 
hole if nature admits it. 
If so, the concept of interior entropy could provide a choice to explain
where the information goes before the black hole shrinks to a remnant 
or appearing white hole\cite{Rovelli:2024sjl}. 
Theoretically, if interior entropy could serve as a new type of entropy which is related to information\cite{Page:1993wv,Rovelli:2017mzl},
the total variation between this type of entropy and 
Bekenstein-Hawking entropy is expected to satisfy the
second law of thermodynamics during Hawking radiation.
  
In addition, quantum field theory predicts that it may exist 
topology defects such as 
cosmic strings\cite{Hindmarsh:1994re,Copeland:2009ga} 
due to occurring of phase 
transition which induced by spontaneous symmetry breaking 
in early universe.
Cosmic strings would survive in the universe in the form of network
which consists of infinite cosmic strings and cosmic string loops.
Also, cosmic strings can give rise to some observable effects. 
For example, when the string network evolves in the spacetime,
cosmic strings will radiate gravitational waves\cite{Vachaspati:1984gt,Allen:1991bk,Damour:2004kw},
perhaps form black holes\cite{Garriga:1992nm},
and could result in gravitational lensing phenomenon\cite{Vilenkin:1984ea}. 
A recent work in \cite{Aurrekoetxea:2023vtp} takes 
numerical relativity method\cite{Helfer:2018qgv,Aurrekoetxea:2020tuw} 
to investigate GW190521 event of gravitational wave 
which obtained from the observation results in the third 
LIGO-Virgo-KAGRA run,
and finds that cosmic string loops that collapse to black holes 
could potentially mimic high-mass binary black hole mergers if
only the ringdown is observed.
Another phenomenon may arise in astrophysics is that
cosmic strings connect to black holes and form topology defect
black holes\cite{Aryal:1986sz}.
It is still unknown that how do cosmic strings
influence black hole's interior entropy as well as 
its relation to 
Bekenstein-Hawking entropy during black hole evaporation 
once this kind of topology defect is considered as all the previously works do not involve this aspect.    

Thus, in this paper we aim to study the influence coming from cosmic string topology defect as well as quantum gravity effects on the evolution relation between interior entropy and Bekenstein-Hawking entropy for black hole with large mass during Hawking radiation and further examine whether the second law of thermodynamics can be satisfied
for the total variation of these two types of entropy.
Here, it is worth noting that some frameworks in e.g. LQG
have suggested quantum modifications of the classical Bekenstein-Hawking entropy-area
law. For example, in \cite{Ghosh:2011fc,Song:2022zit}, the authors obtained the 
modification by studying the properties of thermodynamics of non-rotating isolated 
horizons; In \cite{Lin:2024flv}, the modification is obtained within the effective 
model of loop quantum black hole; In \cite{Long:2024lbd}, the modification investigated
by implementing the calculations of entanglement involving coherent intertwiners.
In this paper, however, we only focus on examining the quantum effects originating from LQG
on the metric directly and ignore all the different quantum modified formulas of classical 
Bekenstein-Hawking entropy.
This paper is organized as follows. In section \ref{sec:2}, we shall give a brief review of a loop quantum-corrected Schwarzschild black hole. Then in section \ref{sec:3}, we are going to construct a model of a loop quantum-corrected black hole pierced by an infinitely straight cosmic string, and derive the evolution relation between its interior entropy and Bekenstein-Hawking entropy during Hawking radiation, and analyze the behavior of total variation of these two types of entropy in this process.
Finally, our summary appears in section \ref{sec:4}. 
Throughout this paper, we use the units so that $G=c=k_B=\hbar=1$ except where specified.

\section {A loop quantum-corrected black hole}\label{sec:2}
In this section, we briefly review a quantum-corrected Schwarzschild black hole\cite{Lewandowski:2022zce} obtained
in LQG using semiclassical approximation method. The metric of this 
semiclassical quantum black hole reads
\begin{equation}
ds^2=-f(r)dt^2+\frac{1}{f(r)}dr^2+r^2d\theta^2+r^2sin^2\theta d\phi^2, \label{a}
\end{equation}
where $f(r) \equiv 1-\frac{2M}{r}+\frac{\alpha M^2}{r^4}$ and $\alpha \equiv 16\sqrt{3}\pi\gamma^3 l_p^2$ is a constant
with dimension of $M^2$
($\gamma$ named Barbero-Immirzi parameter whose value is around $0.2$ 
in LQG, $l_p \equiv \sqrt{\frac{G \hbar}{c^3}}$ is Planck length).
Moreover, as the existence of the quantum modified term $\frac{\alpha M^2}{r^4}$ in the metric function
$f(r)$, there exists a minimal value of $r$ (i.e., the bounce radius $r_b=(\frac{\alpha M}{2})^\frac{1}{3}$)
which leads to the metric \eqref{a} is valid only for $r\ge r_b$.

In order to build a relation between the black hole mass $M$ and the constant $\alpha$,
one can introduce an auxiliary dimensionless variable $\beta$ taking values of an interval $(0,1)$ such that
\begin{equation}
\alpha=\frac{(1-\beta^2)^3}{4\beta^4}M^2, \label{b}
\end{equation}
and
\begin{equation}
f(r)=\frac{Y(r,\beta)W(r,\beta)}{r^4},  \label{c}
\end{equation}
with $W(r,\beta) \equiv r^2+\frac{(1-\beta)M}{\beta}r+\frac{(1+\beta)(1-\beta)^2M^2}{2\beta^2}$ and
$Y(r,\beta) \equiv r^2-\frac{(1+\beta)M}{\beta}r+\frac{(1+\beta)^2(1-\beta)M^2}{2\beta^2}$ respectively.
As can be seen from equation \eqref{b}, the fact that $\alpha$ is actually a constant which makes $\beta \to 0$
equivalent to $M \to 0$ (Minkowski spacetime) and $\beta \to 1$ means $M \to \infty$ (Schwarzschild case).
To ensure there exists black hole in the spacetime, letting $f(r)=0$, it is easy to see that both $Y(r,\beta)$ in the
range $\beta \in (0,\frac{1}{2})$
and $W(r,\beta)$ in the range $\beta \in (0,1)$ have two conjugate imaginary roots. Besides, $Y(r,\beta)$ has two
unequal real roots when $\beta \in (\frac{1}{2},1)$ and two equal real roots as $\beta=\frac{1}{2}$.
Note, for $\beta=\frac{1}{2}$ case, these two horizons of the loop quantum-corrected black hole will degenerate
and form an extremal black hole with minimal mass $M_{min}=\frac{4\sqrt{\alpha}}{3\sqrt{3}}$ which is at zero temperature
\footnote{\cite{Xiang:2013sza,Li:2016yfd} investigate the effects of quantum gravity of spacetime
from the aspects of generalized uncertainty principle.
Depending on effective geometry and the hypothesis that 
quantum gravity effects
would remove the singularity of black hole, \cite{Xiang:2013sza} 
employs an indirect method without
involving the concrete form of quantum gravity and gets the result 
that the final state of a static spherically
symmetric black hole evaporation will be a stable remnant of 
Planck scale with zero temperature.
Indeed, 
the loop quantum-corrected black hole model can be viewed as
a concrete example of equation $(1)$ in \cite{Xiang:2013sza}.}
(the vanishing temperature can be calculated from equation \eqref{wea} 
below by setting $\beta=1/2$).
Therefore, we
only focus on the case of $\beta \in [\frac{1}{2},1)$ in this paper\footnote{Equivalently, if fixing the mass $M$
of the loop quantum-corrected black hole
then $0<\alpha \leq \frac{27M^2}{16}$, wherein $\alpha \to 0$ 
in metric \eqref{a} 
goes to Schwarzschild geometry ($\beta=1$)
and $\alpha \to \frac{27M^2}{16}$ in it corresponds to extremal
black hole ($\beta=\frac{1}{2}$).}.
The radii of outer horizon $r_+$ and inner one $r_-$ can be calculated as
\begin{equation}
r_{+}=\frac{\beta(1+\sqrt{2\beta-1})\sqrt{\alpha}}{\sqrt{(1+\beta)(1-\beta)^3}}, \label{daa}
\end{equation}
\begin{equation}
r_{-}=\frac{\beta(1-\sqrt{2\beta-1})\sqrt{\alpha}}{\sqrt{(1+\beta)(1-\beta)^3}}. \label{dab}
\end{equation}

\begin{figure}[htb]
\centering
\includegraphics[width=0.5\textwidth]{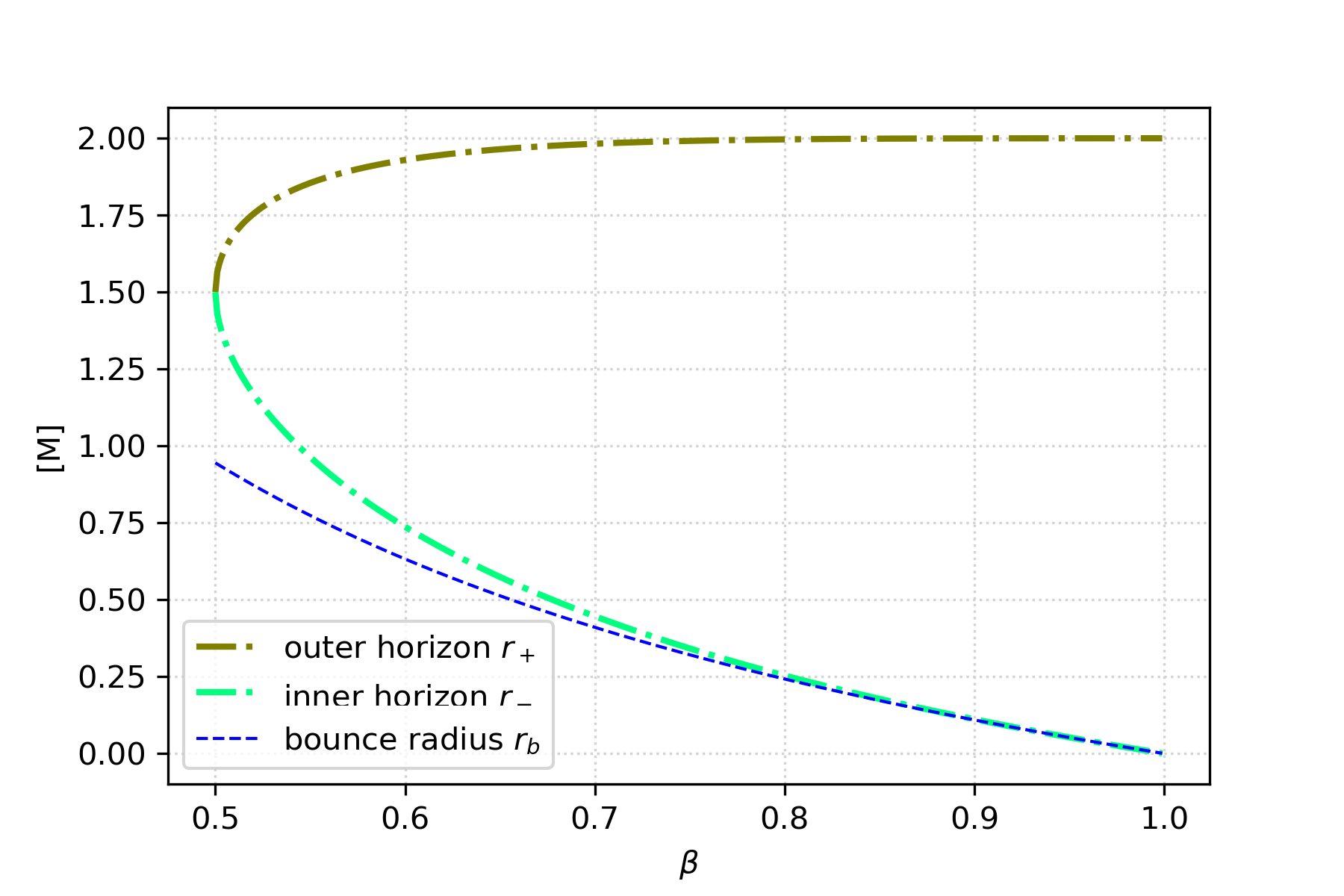}
\caption{Outer horizon $r_+$, inner horizon $r_-$, 
and bounce radius $r_b$
as functions of $\beta$ under fixing 
the loop quantum-corrected black hole's mass $M$.}
\label{figure1}
\end{figure}

To illustrate the influence coming from the quantum effects on the black hole's spacetime structure in a compact
region directly, plugging equation \eqref{b} into equations \eqref{daa} and \eqref{dab} to eliminate $\alpha$, then $r_+$ and $r_-$
can be rewritten as
\begin{equation}
r_{+}=\frac{(1+\beta)M}{2\beta}(1+\sqrt{2\beta-1}), \label{d}
\end{equation}
\begin{equation}
r_{-}=\frac{(1+\beta)M}{2\beta}(1-\sqrt{2\beta-1}). \label{e}
\end{equation}
Also, utilizing equation \eqref{b} the bounce radius then becomes
\begin{equation}
r_{b}=\frac{(1-\beta^2)M}{2\beta^{\frac{4}{3}}}. \label{f}
\end{equation}
We draw function curves of the outer horizon $r_{+}$, 
the inner horizon $r_{-}$,
and the bounce radius $r_b$ in figure \ref{figure1} respectively.
Clearly, the plot depicts that the radius of the outer horizon monotonically increases as $\beta$ growths until approaching
the limit value $2M$ which is consistent with the position of the
event horizon for Schwarzschild case, while the radius of the inner horizon changes
toward the opposite direction till meeting the bounce radius at a limit value $r=0$ (Schwarzschild singularity).

\section {Interior entropy and its relation to Bekenstein-Hawking entropy}\label{sec:3}
First, let an infinitely straight cosmic string pass through a loop quantum-corrected black hole along the axes of $\theta=0$ and $\theta=\pi$, which gives a new metric as\cite{Aryal:1986sz}
\begin{equation}
ds^2=-(1-\frac{2M}{r}+\frac{\alpha M^2}{r^4})dv^2
+(1-\frac{2M}{r}+\frac{\alpha M^2}{r^4})^{-1}dr^2
+r^2d\theta^2+r^2(1-4\mu)^2 \sin^2\theta d\phi^2, \label{g}
\end{equation}
where $\mu$ is the tension (i.e., the mass per unit length) 
of the cosmic string and its value is smaller than unity\cite{Hindmarsh:1994re,Copeland:2009ga}. 
Hence the spacetime corresponding to
the above metric possesses a deficit angle with 
$\delta=8\pi\mu$.
If $\mu$ vanishes, then the metric \eqref{g}
returns to the purely loop quantum-corrected case 
without cosmic string.
Making use of advanced Eddington-Finkelstein coordinate to express the cosmic string black hole metric \eqref{g} which leads to
\begin{equation}
ds^2=-(1-\frac{2M}{r}+\frac{\alpha M^2}{r^4})dv^2+2dv dr+r^2d\theta^2+r^2(1-4\mu)^2 \sin^2\theta d\phi^2. \label{h}
\end{equation}
Unlike the case in \cite{Christodoulou:2014yia} which calculates the interior volume
of a static spherically symmetric black hole, here, we compute the interior volume of a loop
quantum-corrected Schwarzschild black hole with a string under Hawking radiation. So the hole's mass $M$ is a function
of advanced time $v$. The formula of the hole's interior volume is now given by\cite{Ong:2015dja}
\begin{equation}
V_{\Sigma^{\prime}}=\int_{\Omega}(1-4\mu)\sin\theta d{\theta}d{\phi}\int_{0}^{v}\left[\sqrt{-r^4 f(r)}\right]_{max}dv, \label{waa}
\end{equation}
where 
$f(r) \equiv 1-\frac{2M(v)}{r}+\frac{\alpha M^2(v)}{r^4}$. 
It is easy to see that the position of maximal
three-dimensional spacelike hypersurface inside the black hole is
always located at $r_{v}=\frac{3M(v)}{2}$ which is the same as the
classical Schwarzschild case\cite{Christodoulou:2014yia}. 
Further, with a straightforward calculation using equation \eqref{waa} we get the interior volume of the cosmic string black hole, that is
\begin{equation}
V_{\Sigma^\prime}
=(1-4\mu)\pi\int_{0}^{v}M(v)\sqrt{27M^2(v)-16\alpha}dv. \label{wab}
\end{equation}
The variation rate of the above interior volume with respect to time $v$ is given by
\begin{equation}
\frac{dV_{\Sigma^\prime}}{dv}=(1-4\mu)\pi M(v)\sqrt{27M^2(v)-16\alpha}\geq 0. \label{wac}
\end{equation}
Since $M(v)$ decreases as Hawking radiation goes on, equation \eqref{wac} implies that 
the interior volume grows more and more slow.
If the string tension $\mu$ vanishes in equations \eqref{wab} and \eqref{wac}, 
then these two formulas become the interior volume and its
corresponding variation rate for the  loop quantum-corrected black hole, respectively.
The above modified interior volume of equation \eqref{wab} and variation rate of equation \eqref{wac}
would recover the usual results in \cite{Ong:2015dja,Jiang:2020rxx} as $\mu$ vanishes and $\alpha \to 0$.

\begin{figure}[htb]
\centering
\includegraphics[width=0.5\textwidth]{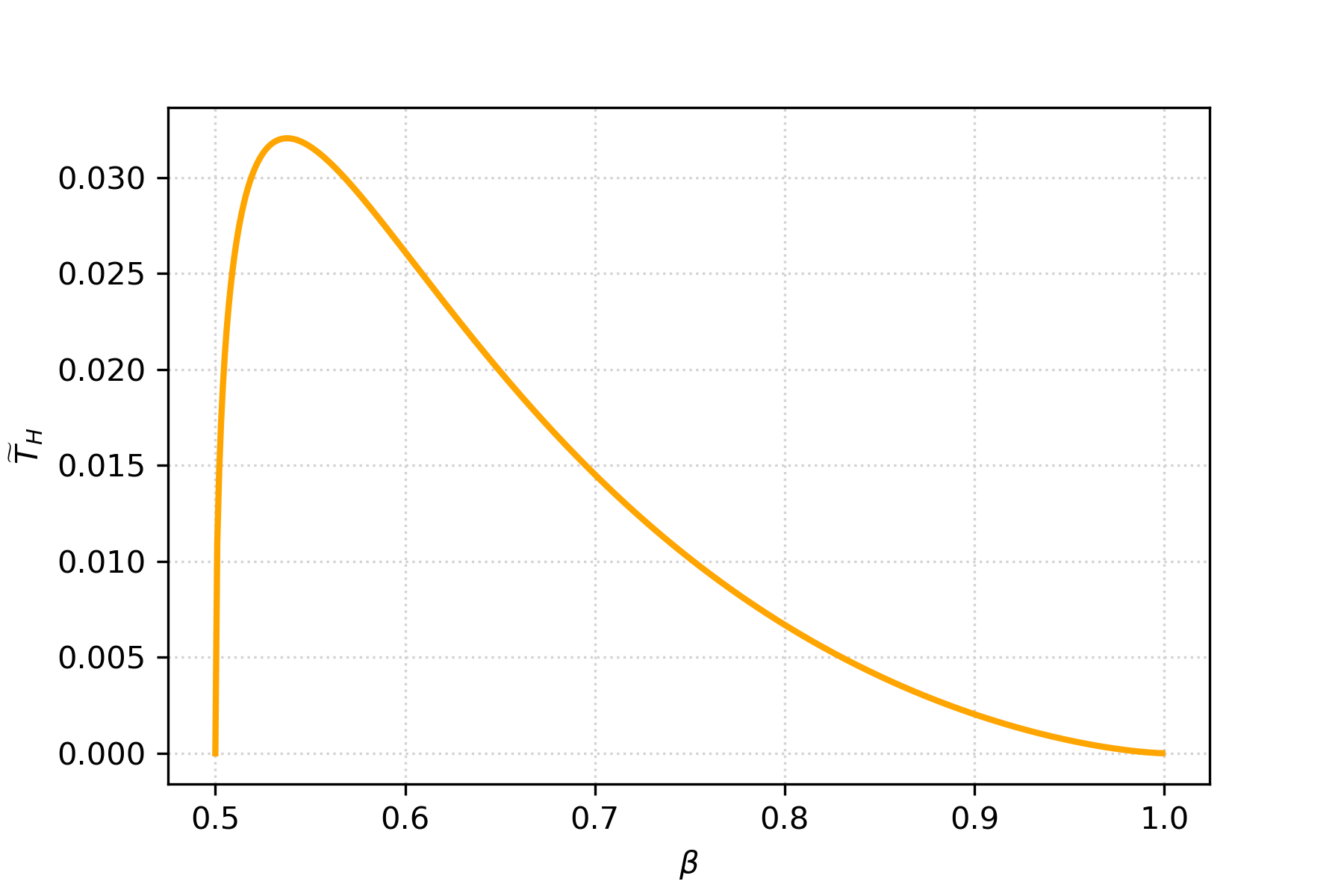}
\caption{$\widetilde T_H$ as a function of $\beta$.}
\label{figure2}
\end{figure}

Let us come to the calculation of Hawking temperature of the horizon surface of the loop quantum-corrected black hole with cosmic string. 
With computing via surface gravity $\kappa$, its Hawking temperature
reads 
\begin{equation}
T_{H}=\frac{\kappa}{2\pi}=\frac{1}{4\pi} \frac{df(r_+)}{dr}=\frac{M}{2\pi r_{+}^2}(1-\frac{2\alpha M}{r_{+}^3}). \label{we}
\end{equation}
Obviously, the cosmic string black hole and the loop quantum-corrected black hole possess the same temperature
and which will also go back to the classical Schwarzschild result $T_H=\frac{1}{8\pi M}$ as $\alpha \to 0$.
With the help of equations \eqref{b} and \eqref{daa}, the above equation \eqref{we} can be rewritten as
\begin{equation}
T_{H}=\frac{(1-\beta)\sqrt{1-\beta}}{\pi \sqrt{\alpha}\sqrt{1+\beta}(1+\sqrt{2\beta-1})^2}[1-\frac{4(1-\beta)^3}{\beta(1+\sqrt{2\beta-1})^3}]\equiv \frac{\widetilde T_H}{\sqrt{\alpha}}, \label{wea}
\end{equation}
where $\widetilde T_H$ is a dimensionless variable.
To describe the variation of Hawking temperature in the process of Hawking radiation, we only need to display the variation
of $\widetilde T_H$ with respect to $\beta$ (see figure \ref{figure2}) since $\frac{1}{\sqrt{\alpha}}$ equals a constant.
Figure \ref{figure2} shows that the loop quantum-corrected 
black holes with and without
cosmic string have maximum temperature when
$\beta\approx 0.538$, and at this moment the masses of the black holes take the value $M\thickapprox 0.966\sqrt{\alpha}$ 
(calculating by using equation \eqref{b}) which are at Planck scale. 
Notice that the temperature of both the black hole interior and its surface of event horizon are the same during Hawking radiation
on condition that the black hole holds large mass. Otherwise, one could not identical the Hawking temperature $T_H$ to the temperature of the massless scalar
field of the black hole interior in an evaporation process because the bulk and surface of the black hole could not in a state of thermal equilibrium.
Therefore, for large mass black hole with and without cosmic string cases, 
both of their interior entropy (according to equation \eqref{ab}) increase as Hawking radiation goes on.

Now, we are going to derive the quantitatively relationship between the interior entropy and Bekenstein-Hawking entropy for a
loop quantum-corrected Schwarzschild black hole with large mass pierced by a cosmic string in the Hawking radiation process. 
Due to Hawking radiation is slow enough for black hole with large mass,
for an infinitesimal time interval, it has been demonstrated in \cite{Wang:2020fgz} that
one can use the initial mass of the black hole 
to calculate its interior volume under leading order approximation.
Meanwhile, the infinitesimal variation of the black hole's Hawking 
temperature $dT_H\sim 0$\cite{Wang:2018txl,Wang:2019dpk,Zhang:2019abv}.
Then the interior volume given in equation \eqref{wab} becomes
\begin{equation}
V_{\Sigma^\prime}\simeq (1-4\mu)\pi M v\sqrt{27M^2-16\alpha}, \label{weg}
\end{equation}
and an infinitesimal variation of interior entropy obeys
\begin{equation}
dS_{\Sigma^\prime}\simeq \frac{\pi^2T_H^3}{45}dV_{\Sigma^\prime}=
\frac{(1-4\mu)\pi^3 T_H^3 M\sqrt{27M^2-16\alpha}}{45}dv. \label{weh}
\end{equation}
Taking into account the Stefan-Boltzmann law
\begin{equation}
\frac{dM}{dv}=-\sigma T_H^4 A, \label{xaa}
\end{equation}
where $\sigma$ being positive 
($\thicksim 10^{-5}$)\cite{Wang:2018dvo,Parker:2009uva} 
is the Stefan-Boltzmann constant and
$A$ is the horizon area of the cosmic string black hole which reads 
\begin{equation}
A=(1-4\mu)r_{+}^2\int_{0}^{\pi}\sin\theta d\theta\int_{0}^{2\pi}d\phi=(1-4\mu)4\pi r_{+}^2. \label{xaaa}
\end{equation} 
In addition, it is observed that $f(r_+)=0$, resulting in 
\begin{equation}
dM=\frac{M(r_{+}^3-2\alpha M)}{r_{+}(r_{+}^3-\alpha M)}dr_{+}. \label{xab}
\end{equation}
Using equations \eqref{xaa} and \eqref{xab}, then equation \eqref{weh} becomes
\begin{equation}
dS_{\Sigma^{\prime}}=-\frac{(1-4\mu)\pi^3 M^2\sqrt{27M^2-16\alpha}(r_+^3-2\alpha M)}{45\sigma T_H Ar_{+}(r_+^3-\alpha M)}dr_+. \label{xac}
\end{equation}

Let us move on to check the validity of classical Bekenstein-Hawking entropy-area law $S_{BH}=A/4$
in the presence of a loop quantum-corrected black hole 
pierced by an infinitely straight cosmic string.
Since the spacetime in this case is not asymptotically flat\cite{Aryal:1986sz},
for an infinity observer, the energy $E$ is not equal to the mass $M$ of the black hole in the situation of black hole with a cosmic string. 
One can use the following equation to calculate the entropy related to the surface of the event horizon of the black hole
\begin{equation}
dS_{BH}=\frac{dE}{T_H}. \label{xaca}
\end{equation}
It is worth pointing out that 
the authors in \cite{Zhang:2024nny} have shown 
that the behavior of the loop quantum-corrected black hole is the same as that of the large-scale behavior of both 
Hayward black hole 
and the regular black hole with $x=1$ and $n=3$ 
in \cite{Ling:2021olm}. 
Moreover, it has also 
been demonstrated in \cite{Bayraktar:2018say} that when a Hayward black hole pierced by an infinitely straight cosmic string, 
the entropy related to the event horizon still equals one quarter 
of the horizon area. 
Thus, the classical Bekenstein-Hawking entropy-area law is valid for the loop quantum-corrected black hole with cosmic string and need not to be modified.

Consider an infinitesimal variation of Bekenstein-Hawking entropy $S_{BH}=A/4$ of the event horizon of the cosmic string black hole
\eqref{h} which gives
\begin{equation}
dS_{BH}=2\pi(1-4\mu)r_+dr_+. \label{xad}
\end{equation}
Then after combining equation \eqref{xac} with the above equation \eqref{xad} one can finally obtain the evolution relation between the interior entropy
and Bekenstein-Hawking entropy for the cosmic string black hole with large mass as follows
\begin{equation}
dS_{\Sigma^\prime}=-\frac{\pi^2 Mr_{+} \sqrt{27M^2-16\alpha}}{180\sigma(1-4\mu)(r_+^3-\alpha M)}dS_{BH}, \label{xae}
\end{equation}
and it is obvious that the evolution relation between this two types of entropy depends on
the string tension $\mu$ and the quantum gravity effects characterized by $\alpha$. 
Also, notice here that the corresponding evolution relation of a
Schwarzschild black hole pierced by a cosmic string can be obtained 
in the absence of quantum effects when
letting $\beta \to 1$ (i.e., $\alpha \to 0$ and $r_{+} \to 2M$) such that
\begin{equation}
dS_{\Sigma^\prime}=-\frac{\sqrt{3}\pi^2}{240\sigma(1-4\mu)}dS_{BH}. \label{xaf}
\end{equation}
If there is no cosmic string connect to the black hole, i.e., $\mu=0$, the above equation becomes nothing but the
corresponding evolution relation for Schwarzschild black hole
evaporation 
\begin{equation}
dS_{\Sigma^\prime}=-\frac{\sqrt{3}\pi^2}{240\sigma}dS_{BH}, \label{xaff}
\end{equation}
which has already been got in \cite{Wang:2018dvo}.

Finally, from equation \eqref{xae}, we get the total variation between the interior entropy and Bekenstein-Hawking entropy in the form
\begin{equation}
\dot{S}_1 \equiv {\dot{S}}_{\Sigma^\prime}+{\dot{S}}_{BH}
=\left[1-\frac{\pi^2Mr_{+}\sqrt{27M^2-16\alpha}}{180\sigma(1-4\mu)(r_+^3-\alpha M)}\right]\dot{S}_{BH}
\equiv \widehat\Gamma(\beta)\dot{S}_{BH}, \label{xag}
\end{equation}
where the dot represents the derivative with respect to the advanced time $v$, and $\widehat\Gamma(\beta)$ is the
proportion function containing the influence of cosmic string.
Again, if setting $\mu=0$, then 
\begin{equation}
\dot{S}_2
=\left[1-\frac{\pi^2Mr_{+}\sqrt{27M^2-16\alpha}}{180\sigma(r_+^3-\alpha M)}\right]\dot{S}_{BH}
\equiv \Gamma(\beta)\dot{S}_{BH}, \label{xah}
\end{equation}
being the total variation of these two types of entropy for the loop quantum-corrected black hole without cosmic string, and  $\Gamma(\beta)$ is the proportion function in this condition.
Remember that, for a black hole with large mass $M$,
the value of $\alpha$ is in the range of $(0,\frac{27M^2}{16}]$  
(actually $\alpha$ is far from $\frac{27M^2}{16}$), 
and $r_+$ belongs to the interval $[\frac{3M}{2},2M)$ (actually
$r_+$ is far from $\frac{3M}{2}$), and the Stefan-Boltzmann constant 
$\sigma\thicksim 10^{-5}$,
it is not difficult to find that $\widehat\Gamma(\beta)<0$ 
and $\Gamma(\beta)<0$.
Since $\dot{S}_{BH}<0$ in the process of Hawking radiation,
one gets that $\dot{S}_1>0$ and $\dot{S}_2>0$ which indicate that the second law of thermodynamics is satisfied in these two situations.

\section {Summary and Discussion}\label{sec:4}
In this paper, we
investigated the interior entropy and its relation to Bekenstein-Hawking entropy for a loop quantum-corrected black hole with large mass passing through an infinitely straight cosmic string in the Hawking radiation process. The obtained results shown that, the present cosmic string not only changed the hole's interior entropy, but also corrected the evolution relation between the interior entropy and Bekenstein-Hawking entropy of the topology defect black hole in comparison with the case of a purely loop quantum-corrected one without cosmic string connect to it, disclosing that the impact of cosmic string
topology defect on the thermodynamic process of black hole evaporation.
Moreover, we found that the total variation of these two types of entropy satisfies the second law of thermodynamics for the loop quantum-corrected black hole no matter it pierced by a cosmic string or not during Hawking radiation,
which further verified the validity that the concept of interior entropy 
as a potential scheme to deal with the black hole information problem. 

As discussion, in this paper we only focus on large mass black hole,
and it will be interesting to consider small mass black hole\cite{Zhang:2021xoa}.  
However, the evaporation process in an infinitesimal time interval for 
small mass black hole is fast, which makes 
the temperature of the interior of the black hole and that of the event 
horizon of it in a non-equilibrium thermodynamics state\cite{Zhang:2019abv}, giving rise to
an obstacle that identifying these two different temperatures 
to establish the evolution relation between interior entropy and 
Bekenstein-Hawking entropy is invalid.

\begin{acknowledgments}
The author would like to thank Cong Zhang 
for useful discussions. This 
work is supported by the National Natural Science Foundation 
of China with Grant Nos. 12075026 and 12361141825.
\end{acknowledgments}

\end{document}